\documentclass[useAMS,usenatbib,usedcolumn,eps,psfig,graphics]{mn2e}
\usepackage{epsfig, graphicx}

\newcommand{\whzsr}{W\,Hz$^{-1}$\,sr$^{-1}$}

\def\sun{\hbox{$\odot$}}
\def\subsun{\mbox{$_{\normalsize\odot}$}}

\def\msol{M$_{\odot}$}
\def\msolyr{M$_{\odot}$ yr$^{-1}$}

\newcommand{\aap}{A\&A}

\newcommand{\aj}{AJ}
\newcommand{\apj}{ApJ}

\newcommand{\araa}{ARA\&A}
\newcommand{\mnras}{MNRAS}
\newcommand{\nat}{Nat}

\def\grtsim{\mathrel{\hbox{\rlap{\hbox{\lower2pt\hbox{$\sim$}}}\raise2pt\hbox{$>$}}}}
\def\lesssim{\mathrel{\hbox{\rlap{\hbox{\lower2pt\hbox{$\sim$}}}\raise2pt\hbox{$<$}}}}
\def\degree{\nobreak\ifmmode{^\circ}\else{$^\circ$}\fi}

\def\size{150 {\rm pc}} % median

\def\rms{${\rm 25~\mu Jy~per~17\times 14~mas^{2}~beam}$}

\def\cfreq{1658.24 {MHz}}

\title[High-redshift obscured quasars: radio emission at sub-kiloparsec scales]{High-redshift obscured quasars: radio emission at sub-kiloparsec scales}
\author[H.-R. Kl\"ockner et al.]{H.-R. Kl\"ockner$^{1}$\thanks{E-mail:hrk@astro.ox.ac.uk},
A. Mart\'\i nez-Sansigre$^{2}$, S. Rawlings$^{1}$, M. A. Garrett$^{3, 4, 5}$\\
$^{1}$Astrophysics, Department of Physics, University of Oxford, Keble Road, Oxford OX1 3RH, United Kingdom\\
$^{2}$Max-Planck-Institut f\"ur Astronomie, K\"onigstuhl-17, Heidelberg, 
  D-69117, Germany \\
$^{3}$Netherlands Institute for Radio Astronomy (ASTRON), Postbox 2, 7990 AA Dwingeloo, The Netherlands\\
$^{4}$Leiden Observatory, University of Leiden, Postbox. 9513, Leiden 2300 RA, The Netherlands\\
$^{5}$Centre for Astrophysics and Supercomputing, Swinburne University of Technology, Hawthorn, Victoria 3122, Australia\\
}
\begin{document}
\topmargin -0.5in
\sloppy
\date{Accepted . Received .}

\pagerange{\pageref{firstpage}--\pageref{lastpage}} \pubyear{2009}

\maketitle

\label{firstpage}

\begin{abstract} 
  The radio properties of 11 obscured `radio-intermediate' quasars at
  redshifts $z\grtsim2$ have been investigated using the European
  Very-Long-Baseline-Interferometry Network (EVN) at 1.66~GHz. A
  sensitivity of $\sim$\rms\, was achieved, and in 7 out of 11 sources
  unresolved radio emission was securely detected.  The detected radio
  emission of each source accounts for $\sim30-100$\% of the total
  source flux density. The physical extent of this emission is
  $\lesssim$~\size, and the derived properties indicate that this
  emission originates from an active galactic nucleus (AGN). The
  missing flux density is difficult to account for by star-formation
  alone, so radio components associated with jets of physical size
  $\grtsim$~\size\, and $\lesssim 40$~kpc are likely to be present in
  most of the sources. Amongst the observed sample steep, flat,
  gigahertz-peaked and compact-steep spectrum sources are all
  present. Hence, as well as extended and compact jets, examples of
  beamed jets are also inferred, suggesting that in these sources, the
  obscuration must be due to dust in the host galaxy, rather than the
  torus invoked by the unified schemes. Comparing the total to core
  ($\lesssim$\size) radio luminosities of this sample with different
  types of AGN suggests that this sample of $z\grtsim$2
  radio-intermediate obscured quasars shows radio properties that are
  more similar to those of the high-radio-luminosity end of the
  low-redshift radio-quiet quasar population than those of FR~I radio
  galaxies. This conclusion may reflect intrinsic differences, but could be
  strongly influenced by the increasing effect of inverse-Compton
  cooling of extended radio jets at high redshift.

\end{abstract}


\section{Introduction}

Active galactic nuclei (AGN) are a class of galaxies that show signs
of non-stellar emission, believed to be associated with processes
connected to accretion onto a supermassive black hole (SMBH, e.g. Rees
1984).  Direct and re-processed radiation arising from the accretion
process itself is responsible for much of the emission of the AGN
between X-rays and the mid-infrared
(e.g. Krolik 1999).  The radio emission, however, is
synchrotron radiation from radio components typically associated with
jets launched from the central region. The jets might be powered by
magnetic fields leaving the accretion disk (Blandford \& Payne 1982)
or by field lines leaving the horizon of the spinning black hole itself
(Blandford \& Znajek 1977).  Star-formation in the host galaxy can
dominate the energy output at far-infrared and sub-millimetre
wavelengths (e.g. Rowan-Robinson 1995) and lead to the emission
of synchrotron emission at radio wavelengths. Such emission is seeded
by cosmic ray electrons accelerated by processes associated with supernovae
(SNe) and their remnants, and synchrotron cooling occurs via
interaction with magnetic fields both local to the sites of SNe and,
following diffusion away from these sites, throughout the AGN host
galaxy (Condon 1992).

The most bolometrically luminous AGN are dubbed quasars (with absolute
blue magnitude $M_{\rm B}<-23$ when they are not obscured), while
lower luminosity AGN are known as Seyferts.  This is mainly a
taxonomically designation, since quasars and Seyferts seem to form a
continous progression in AGN luminosity, most likely controlled by the
accretion rate onto the SMBH. Many of the differences in observed properties
between broad-line (type 1) and narrow-line (type 2) quasars can be
ascribed to orientation-dependent obscuration, perhaps due to a torus
of obscuring material on larger scales than, but approximately
co-aligned with, the accretion disk.  The powerful radio-selected AGN
(radio-loud quasars and radio galaxies) are found to show distinct
structural properties depending on their radio luminosity: the
brightest sources (Fanaroff-Riley 1974, class-II; FR~II) show
double-lobed jets with the brightest emission from hotspots and
head regions at the edge of the lobes, while the less-luminous FR~I
sources have the brightest emission near the base of the jet, with the
surface brightness decreasing with distance from the centre. The range
in radio luminosity is believed to be closely connected to a range in
bulk kinetic powers in the jets in a way that is seemingly connected
to the underlying accretion rate onto the SMBH (e.g. Rawlings \&
Saunders 1991).  When the orientation of the jet is close enough to
the line of sight, Doppler-boosting (beaming) effects can affect the
observed radio emission (e.g. Urry \& Padovani 1995).

Quasars are usually classified as radio-loud (RLQ) or radio-quiet
(RQQ) either from their ratio of radio to optical luminosity
(e.g. Kellerman et al. 1989), with RLQs having a ratio $>$10, or from
their radio luminosities (Miller, Peacock \& Mead 1990), with RLQ
having $L_{5~{\rm GHz}}> 10^{25}$~\whzsr, RQQ having $L_{5~{\rm GHz}}<
10^{24}$~\whzsr. Quasars with radio luminosities in between these
values are classified as radio-intermediate (RIQ), and this radio
luminosity corresponds approximately to that of FR~I radio
galaxies. However, most radio-selected FR~I radio galaxies have
intrinsically weaker bolometric luminosities due to low accretion
rates (e.g. Hine \& Longair 1979).  The large majority of quasars,
$\sim$90\%, are RQQ (Kellermann et~al. 1989, and more recently
Ivezi{\'c} et~al. 2002), and it is still not clear whether there is a
real dichotomy between the RLQ and the RQQ populations (see e.g.
Ivezi{\'c} et~al. 2002; Cirasuolo et~al. 2003).

The physical reason why some quasars are radio-loud is also still unclear,
given that quasars which output similar radiative bolometric luminosities can
show a huge range of radio luminosities, suggesting wildly different jet
powers.  Obsering the Palomar-Green bright quasar sample (BQS) at radio
frequencies, Miller, Rawlings \& Saunders (1993) suggested that jets are
ubiquitous in RQQ as well as RLQ. Miller et~al. (1993) also suggested that RIQ
are intrinsically RQQ, only the line-of-sight of the observer is closely
aligned with the base of the jet.  Relativistic Doppler boosting of the radio
emission makes these RQQ appear brighter to the observer, who classifies them
as RIQ.

The observations of Falcke et al. (1996) and Kukula et al. (1998) supported this
scenario to some extent. The former observed with Very Long Baseline 
Interferometry (VLBI) three flat-spectrum RIQ and found high 
brightness temperatures ($T_{\rm  B}\sim10^{10}$ K) and no extended emission, 
suggesting emission dominated by a Doppler-boosted core.
The latter observed a sample of RQQ from the BQS, which included two RIQ. 
They found these two RIQ to have
flat or inverted spectra, variable fluxes and high values of $T_{\rm B}$,
unlike the RQQ, supporting the beaming scenario. However, they also found
extended emission on scales larger than those observed in RQQ. It must also be
noted that Falcke et al. choose three RIQ that were known to be flat-spectrum,
thus biasing their study, while Kukula et al. only had two RIQ in their sample.
Thus, neither of their results on beaming of RIQ were conclusive.

For a long time there has been a dearth of quasars with
FR~I-structures. However, Blundell \& Rawlings (2001) and Heywood,
Blundell \& Rawlings (2007) have shown this lack of FR~I quasars might
be an artefact of snapshot radio images with relatively long
baselines, which are less sensitive to extended structure with low
surface brightness.  Mart\'\i nez-Sansigre et~al.  (2006b, MS06b)
observed a sample of $z\grtsim2$ obscured quasars with luminosities
comparable to those of RIQ and FR~I radio galaxies, and found most of
them to have steep spectra, thus ruling them out as beamed RQQ.

While the nature of RIQ might still be unclear, it seems certain that
RQQ often possess scaled-down versions of the jets seen in RLQ. RQQ
have mostly, if not always, steep radio spectra, emission extended
over $\grtsim$1kpc scales is sometimes detected (Kukula et al. 1998),
they sometimes show compact cores, and superluminal motion has even
been reported (Blundell \& Beasley 1998; Blundell, Beasley \&
Bicknell 2003).

Kukula et~al. (1998) observed a sample of RQQ at 1.4, 4.8 and 8.4~GHz,
combining data from the VLA in the most extended configuration (A-array), the
most compact configuration (D-array) and complementary published data. They
found that a large fraction of the total emission was recovered, at 4.8 and
8.4~GHz, in ``cores'' with sizes $\lesssim$1~kpc, with a mean recovered fraction
of 60\%. Approximately half of the objects showed evidence of extended
structures on scales $\grtsim$1~kpc, while the other half were unresolved point
sources. The cores were generally found to have steep spectra $S_{\nu}\propto
\nu^{-\alpha}$, with $\alpha=0.7$, although some showed flat spectral indices
($\alpha\leq0.5$).  The inferred brightness temperatures for these cores were
found to be mostly in the range $T_{\rm B}\sim 10^{4}-10^{6}$~K, and thus inconsistent
with emission associated with supernovae (SNe) remnants.

The radio properties of Seyfert galaxies suggest these are the natural
low-luminosity extensions of quasars: most of the sources have steep 
radio spectra, but flatter spectra are sometimes found (Kukula et al. 1995).  
Approximately half of the sources have only compact ($\ll100$~pc) 
emission, with no hint of
extended emission, and the other half show extended emission, sometimes on
scales $\grtsim$1~kpc (Kukula et al. 1995; Gallimore et al. 2006).

This paper presents VLBI observations of a subsample of high-redshift,
radio-intermediate obscured quasars. The original sample was selected
using a combination of data at 24~$\mu$m, 3.6~$\mu$m and 1.4~GHz, and
was used to argue that most of the SMBH growth is obscured by dust
(Mart\'\i nez-Sansigre et~al. 2005).  Optical spectroscopy has shown
that about half of the sources do not show any rest-frame ultraviolet
emission lines, the other half showed narrow emission lines only
(Mart\'\i nez-Sansigre et~al. 2006a, hereafter MS06a).  Combining
optical and mid-infrared spectroscopy, 18 out of 21 of the sources in
the sample have spectroscopic redshifts in the range $1.6 \leq z \leq
4.2$ (MS06a; Mart\'\i nez-Sansigre et~al. 2008, hereafter MS08). It has
been argued that the obscured quasars come in two flavours: classic
type 2 objects in which the quasar nucleus is obscured by a torus and
``host-obscured'' objects in which the quasar nucleus is obscured by
dust distributed on larger scales within the host galaxy, and possibly
associated with ongoing star formation.  The inferred radio
luminosities are similar to both RIQ and FR~I radio sources (MS06b).

Multiple radio frequencies showed most of the sources have very steep
spectra $\alpha\sim1$ (MS06b). This is consistent with lobe emission
with the same time-steady electron injection spectrum as
lower-redshift radio galaxies [electron number densities $N(E) \propto
E^{-p}$ with $p\sim2.5$], provided that the increased inverse-Compton
losses against the Cosmic Microwave Backgound (CMB) at high redshift
cause the electron energy spectrum to steepen [$N(E)\propto
E^{-(p+1)}$ instead of $N(E)\propto E^{-p}$] at the rest-frame
frequencies corresponding to those observed. The observed flux density
is then expected to be $S_{\nu} \propto \nu^{-{1\over 2}(p+1-1)}$
instead of $\propto \nu^{-{1\over 2}(p-1)}$, and will therefore have a
spectral index $\alpha\sim1.25$ rather than $\sim 0.75$ as observed
for lower-redshift sources (see e.g. Kardashev 1962).

Indeed, from `minimum energy' arguments and assuming $L_{\rm 1.4~GHz}=
10^{24}$ W Hz$^{-1}$ sr$^{-1}$ with a characteristic size of 30 kpc
and age of 10~Myr, a $B$-field of 1-2~nT was inferred by MS06b,
comparable to the field inferred in FR~I radio galaxies (e.g. Laing et
al. 2006). The combined effect of synchrotron ageing in this magnetic
field and inverse-Compton losses against the CMB at $z\sim2$ could
then explain the steep spectra.  Comparison of 1.4~GHz data at 5- and
14-arcsec resolution, showed no significant difference in flux density
(at the $\geq$2--3$\sigma$ level) for all but one source (AMS04, see
MS06a).  This suggests that at rest-frame $\grtsim$5~GHz, there is a
negligible amount of flux density on scales $\grtsim$40~kpc.  However,
an older population of electrons at these large distances cannot be
ruled out by the relatively high (rest-frame) frequency data used by
MS06a and MS06b. Mapping such emission would require low-frequency
observations with an array sensitive to very extended emission.

Since for the MS06b sample the rest-frame $\sim$5~GHz emission is
relatively compact ($\lesssim$5 arcsec corresponding to
$\lesssim$40~kpc), any further spatial studies of these sources
require very long baselines. Such VLBI studies can determine whether
the steep-spectrum lobes have characteristic sizes $\sim$kpc or
$\ll$kpc and can also determine whether flat-spectrum cores are
present. While most of the sources showed very steep spectra, some of
the sources showed gigahertz-peaked spectra, and surprisingly for
obscured quasars, some sources showed flat spectra suggesting face-on
beamed jets (MS06b). However, all the flat-spectrum sources were found
to have blank optical spectra. This is consistent with such objects
being obscured by dust on galactic scales which also hides the
narrow-line region, instead of being obscured by the torus. Such
flat-spectrum sources with no optical lines are likely examples of
``host-obscured'' quasars.  Observations with VLBI are again necessary
to place further constraints on any beamed components, which are
expected to show compact, high-$T_{\rm B}$ emission.

This paper presents radio observations of a subsample of 11 out of 21
sources in the aforementioned sample of high-redshift
radio-intermediate obscured quasars, using the European VLBI
Network (EVN).  The subsample includes all sources with spectroscopic
or crude photometric redshifts z$\geq$2 (as published in Mart\'\i
nez-Sansigre et~al. 2005; MS06a). More recent optical and mid-infrared
spectra have confirmed 10 out of the 11 sources presented here as obscured
quasars, with spectroscopic redshifts in the range $1.8 \leq z \leq
4.2$ (MS06a; MS08).

The observations and data reduction are presented in Section~\ref{sec:obs}.
Section~\ref{sec:res} summarises the results of the observations, while
Section~\ref{sec:dis} discusses the physical implication of the detected
emission and the flux missed by the EVN.  The conclusions and summary are
presented in Section~\ref{sec:conc}.  Throughout this paper a $\Lambda$CDM
cosmology is assumed with the following parameters: ${\rm h} ={\rm H_0}/(\,100
\,{\rm km\,s^{-1}}\,{\rm Mpc^{-1}}) = 0.7$, $\Omega_m = 0.3$, $\Omega_{\Lambda}
= 0.3$.

\section{Observations and data reduction}\label{sec:obs}

\begin{figure}
\includegraphics[bb=20 20 580 580,clip,angle=0,width=8cm]{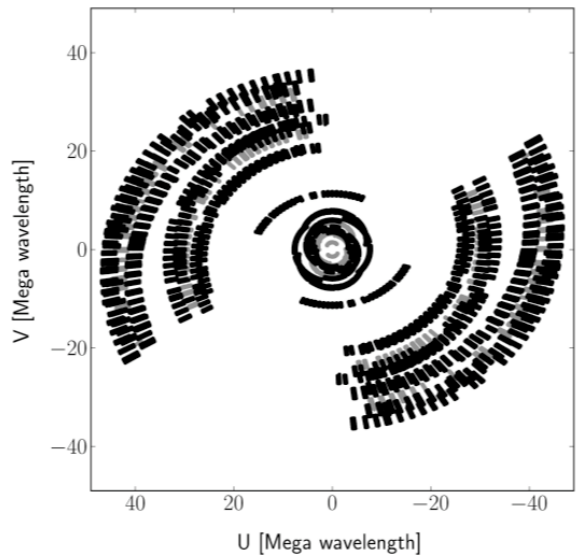}
\caption{Example of the UV-coverage of
  the phase-calibrator source J1722+5856, observed 29th October
  2005 (date A). Gray points display the UV coverage lost by excluding the WSRT
  array. The radial extend of the individual scans are due to the
  large frequency coverage of 128~MHz.}
\label{fig:uvcover}
\end{figure}

\begin{table*}
\begin{center}
\begin{tabular}{cccccl}
\hline
\hline
calibrator    & & amplitude on individual  run  [mJy] & &  mean $\pm$ rms [mJy] & calibrated with \\
 & A & B & C & \\ \hline
J1722$+$5856 & 136.90 $\pm$ 2.14 & 141.48 $\pm$ 0.81 & 130.76 $\pm$ 1.91 & 136.38 $\pm$ 5.37 & J1722$+$5856 \\
J1722$+$6105 & 174.70 $\pm$ 9.13 & 152.91 $\pm$ 21.9 & 174.31 $\pm$ 34.7 & 167.31 $\pm$ 12.5 & J1722$+$5856 \\
J1722$+$6105 & 167.02 $\pm$ 5.33 & 160.27 $\pm$ 1.40 & 170.38 $\pm$ 3.50 & 165.89 $\pm$ 5.15 & J1722$+$6105 \\
\hline
\hline
\end{tabular}
\caption{Calibrator flux densities. J1722+5856 was used as a phase
  reference source for all the three observation runs. Source
  1722+6105 was used to test the phase-referencing technique and
  quantify uncertainties in amplitude. Using J1722+6105 calibrated
  with J1722+5856, an estimate of the night-to-night variation in
  amplitude is obtained, with an uncertainty of 7\%
  (167.31$\pm$12.5). Comparing the amplitudes of J1722+6105 calibrated
  using J1722+5856 and with itself, the uncertainty in amplitude due
  to phase referencing is estimated to be 5\% on any individual run
  (e.g. compare 152.91 and 160.27 on date B). Adding these in quadrature the
  total amplitude uncertainty is therefore estimated to be 9\%. }
\label{tab:calobs}
\end{center}
\end{table*}

\begin{table*}
\begin{center}
\begin{tabular}{ccccccccc}
\hline
\hline
source    & z & optical & radio & peak flux density [mJy] & rms [mJy] & significance [sigma] & $\Delta$ra [mas]& $\Delta$dec [mas] \\ \hline
AMS01$^B$ & 2     & B  & SSS & 165.03  & 33.67  &  4.9  & -672  & -203 \\
AMS03$^B$ & 2.698 & NL & SSS & 1118.00  & 45.24  &  24.7  & 98  & 35 \\
AMS05$^B$ & 2.850 & NL & GPS & 856.30  & 36.72  &  23.3  & -7  & -181 \\
AMS06$^B$ & 1.8   & B  & SSS & 147.79  & 31.23  &  4.7  & 1327  & 881 \\
AMS09$^C$ & 2.1   & B  & SSS & 249.3  & 32.6  &  7.6  & -185  & 36 \\
AMS12$^A$ & 2.767 & NL & SSS & 222.1  & 41.5  &  5.4  & -70   & -123 \\
AMS15$^A$ & 2.1   & B  & FSS & 198.6  & 28.2  &  7.0  & 213   & 140 \\
AMS16$^A$ & 4.169 & NL & GPS & 163.9  & 29.3  &  5.6  & -55   & 1432 \\
AMS17$^A$ & 3.137 & NL & SSS & 151.3  & 29.3  &  5.2  & 967   & -843 \\
AMS19$^C$ & 2.3   & B  & FSS & 599.4  & 38.1  &  15.7 & 10    & 34 \\
AMS21$^C$ & 1.8   & B  & SSS & 221.7  & 30.0  &  7.4  &  90   & 86 \\
\hline
\hline
\end{tabular}
\caption{The target source names are labelled with the observing dates
  A, B, or C. It was not possible to optain spectroscopic redshift for
  AMS01, so a fiducial value of 2 is assumed. For all other sources
  spectroscopic redshifts are quoted. Redshifts with three decimal are
  from optical spectroscopy, while those with one decimal place are
  from mid-infrared spectroscopy (MS06a; MS08). The redshift for AMS05
  comes from Smith et al. (2009). The third column summarizes the
  optical spectroscopic properties: B stands for blank spectrum, NL
  for narrow lines.  The fourth column summarizes the low resolution
  radio spectral indices: FSS stands for flat-spectrum source, GPS for
  gigahertz-peaked source, and SSS stands for steep-spectrum
  source. Target peak flux densities per beam. Amplitude measurements
  of the individual target sources are based on the dirty images. The
  offsets $\Delta$ra and $\Delta$dec are with respect to the B-array
  (1.4~GHz) VLA positions of Condon et al. (2003).}
\label{tab:dirtso}
\end{center}
\end{table*}

\begin{table*}
\begin{center}
\begin{tabular}{lclllccc|cc}
\hline
\hline
Name & Date$^{a}$ &  RA$^{b}$ & Dec &  $S_{peak}$$^{c}$    & $S_{int}$$^{c}$ & $\theta_M$x$\theta_m$$^{d}$ & Distance$^{e}$ & 
$\alpha^{1.4}_{0.6}$$^{f}$ & $\alpha^{4.9}_{1.4}$$^{f}$ \\
     &      & [J2000] & [J2000] & [$\mu$Jy/beam]& [$\mu$Jy] & [mas$^{2}$]           & [deg] & 
& \\
\hline
AMS01&$B$ &               &                &  $<$139.2 /   27.9 &  $<$136.2  / 47.6        &       & 1.242 & 1.0$\pm$0.2 & 0.9$\pm$0.6 \\  
AMS03&$B$ & 17 13 40.2037 & +59 27 45.795  & 1139.4$\pm$44.6 & 1238.3$\pm$81.6            & 18x13  & 1.251 & 1.2$\pm$0.2 & 1.4$\pm$0.3 \\ 
AMS05&$B$ & 17 13 42.7650 & +59 39 20.039  &  923.3$\pm$46.6 &  992.9$\pm$84.8             & 18x14 & 1.333 &  0.2$\pm$0.2 & 0.7$\pm$0.2 \\
AMS06&$B$ &               &                &  $<$128.7 /   25.5 &  $<$125.3  / 43.3      &         & 1.227 & 1.1$\pm$0.2 & 0.7$\pm$0.5 \\  
AMS09&$C$ & 17 14 34.8448 & +58 56 46.467  &  233.9$\pm$33.0 &  326.9$\pm$71.8             & 24x16 & 1.036 & 1.2$\pm$0.2 & 0.7$\pm$0.5 \\
AMS12&$A$ & 17 18 22.6407 & +59 01 54.135  &  210.2$\pm$27.1 &  240.7$\pm$51.3             & 19x14 & 0.552 & 1.2$\pm$0.2 & 0.9$\pm$0.5 \\  
AMS15&$A$ & 17 18 56.9549 & +59 03 25.091  &  201.7$\pm$24.5 &  202.4$\pm$42.5             & 25x23 & 0.485 & 0.3$\pm$0.2 & 0.2$\pm$0.3 \\  
AMS16&$A$ &               &                &  $<$158.8  /  20.3 &  $<$210.5   / 42.5       &       & 0.407 & $<$-1.0 & $>$1.02 \\  
AMS17&$A$ &               &                &  $<$154.1  /   22.5 &  $<$139.4  / 36.5       &       & 0.499 & 1.0$\pm$0.2 & 0.7$\pm$0.4 \\  
AMS19&$C$ & 17 20 48.0000 & +59 43 20.687  &  642.8$\pm$69.9 &  789.5$\pm$139.             & 21x16 & 0.816 & 0.2$\pm$0.1 & 0.4$\pm$0.2 \\  
AMS21&$C$ & 17 21 20.0974 & +59 03 48.644  &  231.8$\pm$30.5 &  258.6$\pm$56.8             & 18x14 & 0.206 & 1.0$\pm$0.2 & 0.9$\pm$0.7 \\
\hline
\hline
\end{tabular}
\caption{ \noindent Radio properties of obscured radio intermediate
  quasars observed with the EVN. Each source has been observed for
  $\sim$1~hour with a bandwidth of 128~MHz at a central frequency of
  \cfreq . $^{a}$Dates $A$ and $B$ correspond to the 29 and 30 October
  2005, respectively, and date $C$ to the 1st November 2005. $^{b}$The
  estimated positions of the EVN radio emission. Comparing the EVN
  positions with the VLA-B positions at 4.9~GHz the maximum offset is
  255~mas. $^{c}$The peak and integrated flux densities and their errors are
  based on a Gaussian model fit to individual sources. For consistency
  the measurements have been determine for all sources in the same
  manner. For sources with a significance $\geq$6$\sigma$ and AMS12 are
  provided, whereas for the non-detection the flux densities and
  errors are given as upper limits. $^{d}$$\theta_M$ and $\theta_m$ are
  the mean semi-major and semi-minor axis of the fitted Gaussian,
  respectively, whereas the array convolved beam is
  17$\times$14~mas$^2$. $^{e}$The distance is the radial separation of the
  observed sources to the phase-reference source J1722+5856. Distances
  of the undetected sources are based on their VLA-B positions (MS06b). $^{f}$The spectral indices
  $\alpha$ and errors between 610~MHz and 1.4~GHz [GMRT, VLA], and between 1.4 and
  4.9~GHz are based on the measurements from MS06b [VLA].}
\label{tab:obs}
\end{center}
\end{table*}

EVN observations at 1.66~GHz were performed on the 29th, 30th October and 1st
November 2005, hereafter dates $A$, $B$ and $C$. The telescopes used during
these observations form an array with baselines ranging of 266~km to 8476~km
[Effelsberg, Onsala (85~ft), Jodrell Bank (Lovell), Medicina, Torun, Urumqi,
Shanghai and WSRT (phased array)]. The WSRT data on the target sources were
lost due to technical problems. This results in an array having a shortest
baseline of $\sim$637~km which corresponds to a spatial sensitivity of
$\lesssim$62~mas. Flux which is extended over scales larger than $\sim$62~mas is
invisible to the interferometer.

The observations used the Mark5A recording system (1024~Mbit/s) with
2-bit sampling, in 2 polarizations and 2s integration time.  The high
data rate capability allows the simultaneous observation of 8
sub-bands. Each of the sub-bands has 32 channels with a bandwidth of
16~MHz. Such a large bandwidth could not be handled at the WSRT and the
Medicina observatory and therefore one sub-band was lost from these
sites. The total frequency coverage is from 1594.99 MHz to 1704.49 MHz
which results in a central frequency of 1658.24 MHz. The large
fractional bandwidth $\sim$8\% significantly increases the UV
coverage. As an example, the UV-coverage of the phase calibrator
source is shown in Figure~\ref{fig:uvcover}.  With an on-source
integration time of $\sim$1~hour, the expected sensitivity in a
naturally weighted image is $\sim$20~$\mu$Jy.

Due to the faintness of the actual targets, the observations made use
of the phase-reference technique, observing a target and a
phase-calibrator source within a cycle of 13 minutes (target source
10~minutes and J1722+5856 for 3~minutes). The target sources are
between 0.2 and 1.5 degrees apart from the phase-calibrator
source. Additional sources were observed to find initial fringes
(J2005+7702, 3C345) and to cross check the phase-referencing technique
(J1722+6105, J171156.0+590639).

\begin{figure}
\includegraphics[angle=0,width=9.0cm]{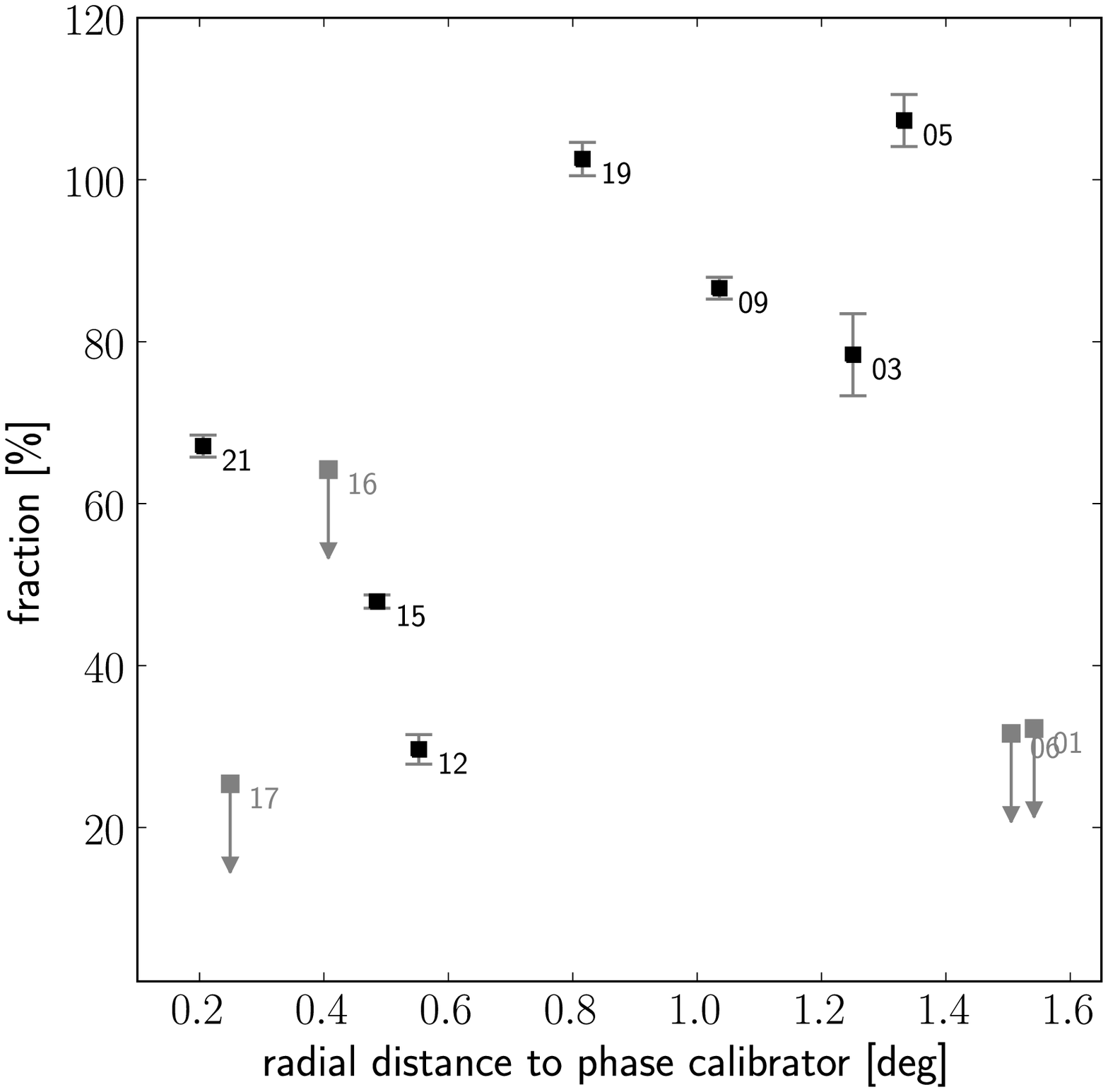}
%\vspace{11pc}
\caption{Recovered radio emission in percent versus the radial
  distance to the phase-calibrator source J1722+5856. The black 
  squares indicate the detected sources. The gray squares
  and arrows indicate the 6$\sigma$ upper limits of the
  undetected sources. The median value is 64 \%. No significant
  correlation is found (null hypothesis of no correlation has a
  probability of 52\% and cannot be rejected [using the ASURV
  software; Isobe, Feigelson \& Nelson, 1986; Isobe \& Feigelson,
  1990]). This supports our contention (see Section~\ref{sec:obs})
  that there is no systematic loss in mapped EVN target flux density
  with separation from the phase calibrator as might be expected if
  time variations in ionospheric phase were the cause of systematic
  reductions in the EVN flux density of the sources. }
\label{fig:perc}
\end{figure}

After observations the data were correlated at the Joint Institute for
VLBI in Europe [JIVE]. The positions of the target sources were
determined on the basis of VLA-B array observations at 5~arcsec
resolution (FWHM). The integration time (2s) and the number of
channels assure that the usable field of view (FoV) of the EVN
observation is larger than the synthesised VLA beam.  Furthermore,
this setup ensures that the expected loss in amplitude due to
time-smearing will only reach $\sim$10\% at a distance of 16.7~arcsec
from the pointing centre\footnote{Based on the EVN sensitivity
  calculator at: http://www.evlbi.org and assuming a point source
  response.}. Similarly, the loss of amplitude due to bandwidth
smearing will only reach $\sim$10\% at 9.9 arcsec. An additional
effect that could cause a loss in flux density is the variation in the
phases caused by the ionosphere. Plotting the fraction of the detected
radio emission versus the angular distance to the phase centre does
not show any hints of a negative correlation (Fig.~\ref{fig:perc}) and
therefore any such
flux density losses can probably be neglected.\\

\begin{figure}
\vspace{0.8cm}
\includegraphics[bb=3 128 532 657, clip,angle=0,width=8cm]{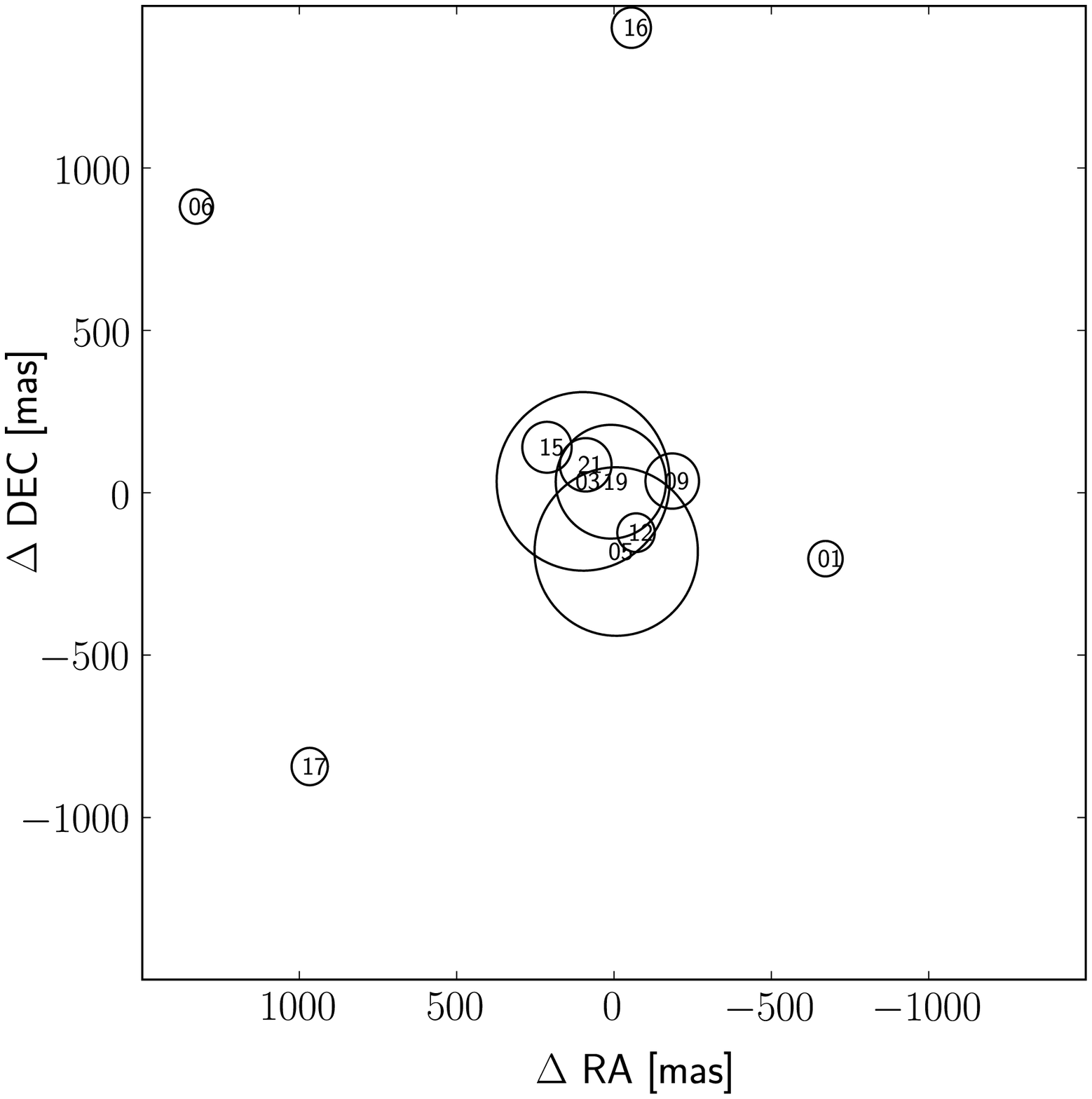}
\caption{Displacement in ra and dec of the peak flux densities in the
  dirty images compared with the VLA positions from Condon et
  al. (2003). The radii of the circles are proportional to the
  significance of the peak flux density compared to the noise in the
  dirty map.}
\label{fig:souredec}
\end{figure}

Data reduction and analysis were performed using AIPS and
ParselTongue \citep{aips,PT2006}. To make sure that all the observing
runs were treated in equal manner, a ParselTongue script was
developed to calibrate the individual observing runs.

Prior to the calibration the datasets were inspected for bad data. All
datasets show narrow band radio interference (RFI) some of which is
variable with time. In all the observation runs the RFI was detected
in the same channels and sub bands and could be flagged in a similar
way using the AIPS task {\tt UVFLG}. Additional flagging was applied using the
flagging tables provided by the VLBI scheduling program SCHED and by
the AIPS task {\tt QUACK}. To account for instrumental polarisation
effects a parallactic angle correction has been applied by {\tt
  CLCOR}. The observational data were a-priori gain (amplitude)
calibrated by using the system temperatures ({\it T$_{sys}$}) measured
at each individual telescope ({\tt ANTAB}). Prior dispersive delay
corrections are applied by the AIPS task {\tt TECOR} using the
ionosphere electron content obtained from the Jet Propulsion
Laboratory (see also the explanation file at the {\tt TECOR} task in
AIPS). The sub-band phase offsets have been corrected using the task
{\tt FRING} on the fringe-finding calibrator J2205+7752 for the
observations A, B and 3C~345 for C. The residual delays and
fringe-rates were then calibrated by initial fringe finding on the
phase-reference source J1722+5856 (using the AIPS task {\tt FRING};
for a full explanation on fringe finding see Cotton 1995).

\noindent The bandpass of the system has been calibrated by using the
phase-reference-source J1722+5856 ({\tt CPASS}). After applying the
amplitude and phase calibration solution and the bandpass the data
show significant offsets in the amplitude over the entire frequency
space, whereas the phase stays constant within 1~degree. Such offsets
in amplitude can be explained by faulty {\it T$_{sys}$} measurements
caused, e.g. by RFI, which are present at most observatories and
individual sub-bands. The amplitude offsets could not be corrected by
a full self-calibration procedure on the phase calibrator itself ({\tt
  CALIB}). Instead the error has been treated as a baseline based
error and the task {\tt BLCAL} has been used to determine a single
solution of the phase and the amplitude for the entire
observation. After applying the baseline corrections, amplitude
offsets between the individual sub-bands were no longer present.

The final UV-dataset has been produced by applying the calibration
correction of J1722+5856 in phase, in amplitude, bandpass, and
baseline excluding 4 channels on each edge of each sub-band. Reducing
the total bandwidth in frequency to 96~MHz and the theoretical image
sensitivity to $\sim$22.5~$\mu$Jy/beam. The final flagging stage of the
UV-dataset was based on the average amplitude using the AIPS task {\tt
  VPLOT}.

In order to test the reliability of the phase-calibration technique,
two test sources (J1722+6105 and J171156.0+590639) were observed and
calibrated in an identical manner to the target sources. Furthermore,
to determine the error on the amplitude measurements between the
different observing runs, the VLBI calibrator J1722+6105 was
additionally calibrated by itself, repeating the same 
steps as J1722+5856 after the sub-band phase offset calibration. The
resulting radio flux density of the calibrator J1722+5856 and the
reference calibrator J1722+6105 using both methods are shown in
Table~\ref{tab:calobs}.

The angular separation between the phase-reference source J1722+5856
and J1722+6122 is 2.16~deg. The amplitude of J1722+6105 resulting from
both types of calibration (i.e. using J1722+5856 and using J1722+6105
itself) have a largest intraday difference of 5\% on night B.  This
yields an estimate on the maximal uncertainty on the amplitude in
using the phase-reference approach in calibrating targets less than
2~deg away from the phase-reference source.

The amplitudes of the phase reference source J1722+5856 shows a
night-to-night variation of about 4\%. The test phase calibrator
J1722+6105, calibrated in a similar way to the targets sources shows during the
3 observing runs an amplitude uncertainty of about 7\%. Whereas
calibrating the calibrator with itself, it shows a night-to-night
variation of about 3\%. Combining the estimates of the uncertainties
from phase referencing and from night-to-night variation, 5\% and 7\%,
yields an estimate of the total uncertainty in amplitude of $\sim$9\%.

After calibration, the position of the phase-calibrator J1722+5856 is
consistent with its position in the VLBI catalogue C-VCS1
\citep{vcs1}. The position of the test phase-calibrator J1722+6105 after 
applying the solutions of the phase-reference source J1722+5856 is
accurate within 3~mas with respect to the C-VCS1 catalogue.

The percentages of the flux densities recovered at EVN resolution (see
Section~\ref{sec:res}) for each target source are shown in
Figure~\ref{fig:perc}. There is no evidence of any systematic
reduction of the recovered flux density with distance from the phase
calibrator, i.e. a negative correlation in Figure~\ref{fig:perc}.

\begin{figure*}
\hspace{-0.5cm}
\vspace{0.1cm}
\includegraphics[bb=20 20 530 700, clip,angle=0,width=16cm,angle=0]{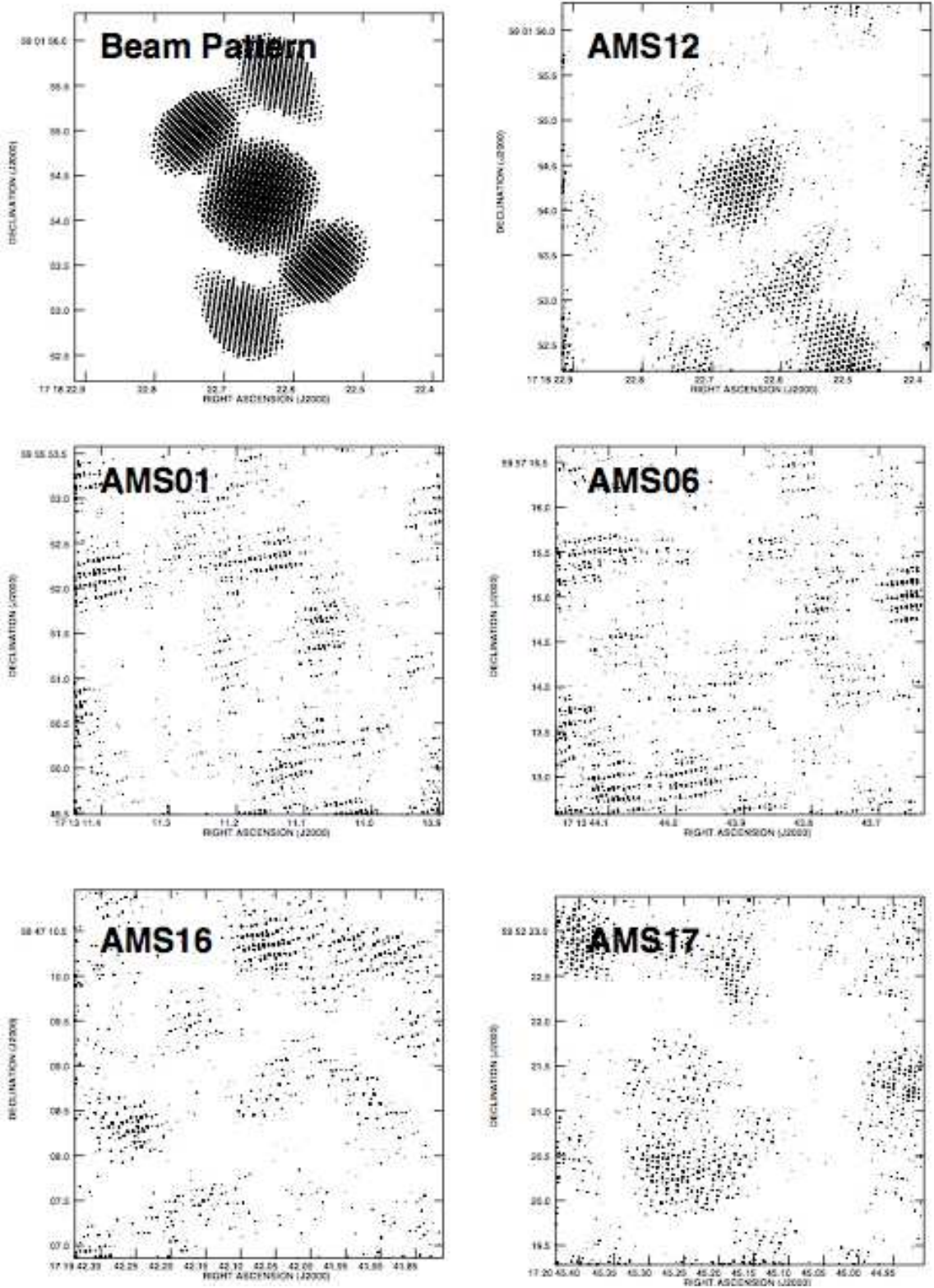}
\vspace{0.2cm}
\caption{Contour plots covering approximately 4.1$\times$4.1 arcsec$^2$
  of the beam pattern (for the observations of AMS12) and the dirty
  images to evaluate detections for sources with peak flux densities
  below 6$\sigma$. The reference image of the beam has been displayed
  in the top left corner for which the contours represent 10, 20, 30,
  and 35\% of the maximum. Only positive contours are shown and these
  peaks and the valleys between these form a grain pattern of
  characteristic size of the order of the resolution set by the
  longest baseline. The image shows the characteristic beam and
  side-lobe levels. For the targets the contours represent 2, 4, 5,
  6, 7 $\sigma$, where $\sigma$ is the rms of the image as quoted in
  Table~\ref{tab:calobs}. The source AMS12 shows the characteristic
  dirty beam and is considered to be a reliable detection.}
\label{fig:dirtyimages}
\end{figure*}

The imaging procedure was performed in two steps (using the AIPS task {\tt
  IMAGR}). First, a dirty image, with robust~=~5 (similar to natural) weighting
applied, was produced. The field-of-view (FoV) of these dirty
images is approximately 4.1$\times$4.1 arcsec$^{2}$, covering
practically the entire synthesised beam of the low-resolution VLA
images.

For each image, the noise and the highest peak flux density have been
determined. For each source these values are shown in
Table~\ref{tab:dirtso} and the significance of the peak flux densities is
represented in Figure~\ref{fig:souredec}. Five sources show peak flux
densities below 6~$\sigma$ and are shown in
Figure~\ref{fig:dirtyimages}. Of these sources, 4 are displaced more
than 500~mas from the VLA positions (which corresponds to the
uncertainty in the Condon et al. 2003 VLA B-array positions at
1.4~GHz).

An example of the synthesized beam pattern is shown in
Figure~\ref{fig:dirtyimages}. Due to the imaging weighting scheme, the
beam pattern produces a large noise floor effecting the determination
of the noise.  Real detected point sources should show the same beam
pattern, although it might be slightly displaced from the VLA
positions (the centre of each image). Visual inspection of
Figure~\ref{fig:dirtyimages} suggests AMS12 is a safe detection while
AMS01 and AMS06 are non-detections. AMS16 and AMS17 show some flux
above 5$\sigma$, displaced from the VLA positions. From the images,
the significance of the peak flux, and the displacement, these two
sources are not considered safe detections. Therefore 6$\sigma$ limits
are quoted for the non-detections.

\begin{figure*}
\vspace{0.1cm}
\includegraphics[bb=15 15 420 750, clip,width=12cm]{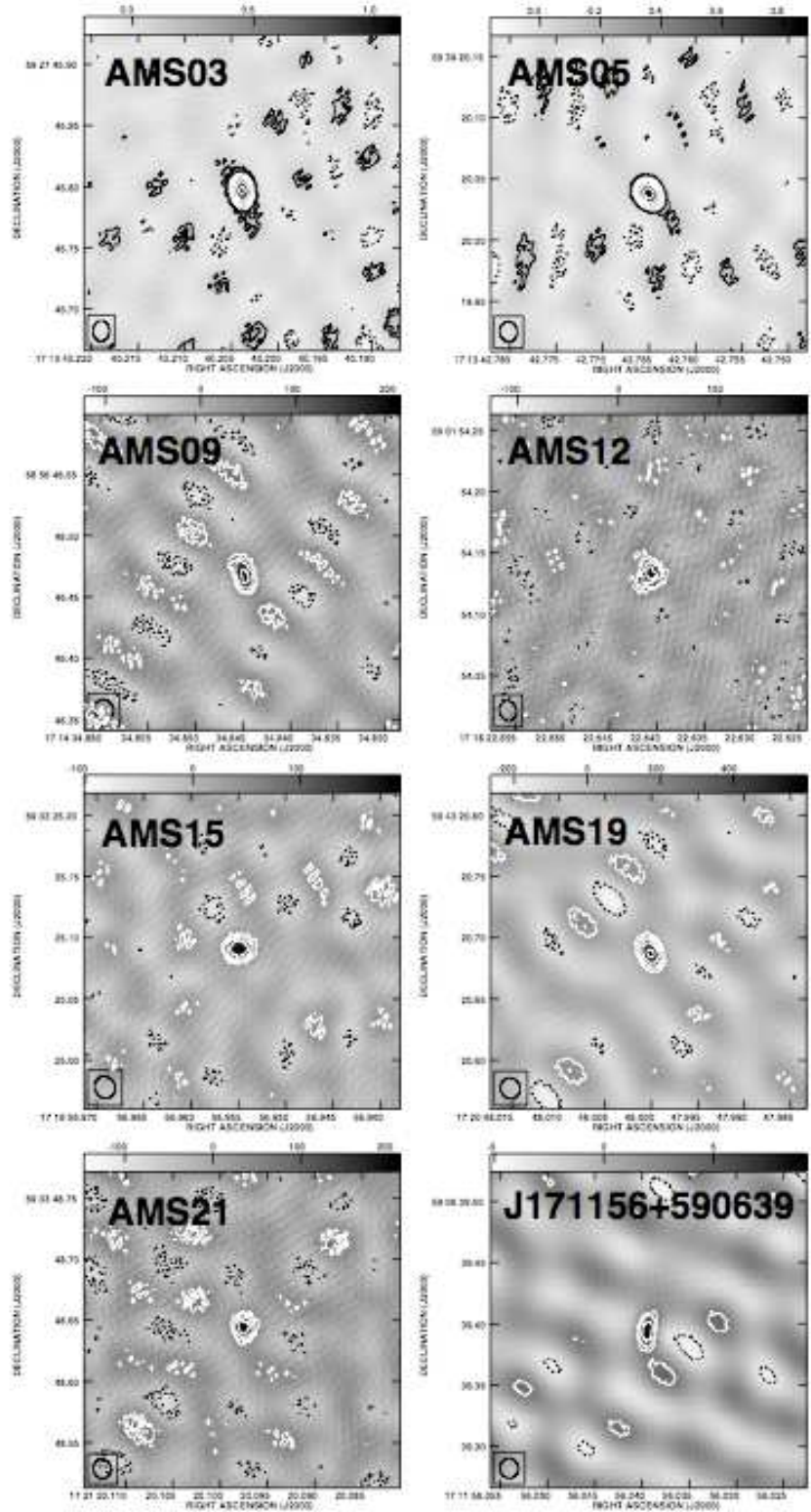}
\vspace{0.0cm}
\caption{EVN cleaned images covering approximately 260$\times$260
  mas$^2$ of the radio emission (Stokes I) at \cfreq\, of the 7
  securely detected high-redshift obscured quasars. The boxed ellipses
  show the Gaussian restoring beam at FWHM.  The contours are at
  30~$\mu$Jy $\times$ 2$^{\frac{n}{2}}$ (n =-2, 2, 3, etc.). The image
  at the bottom right displays the radio emission of the comparison
  source J171156.0+590639 (Wrobel et al. 2004) with contours of
  1.78~mJy $\times$ 2$^{\frac{n}{2}}$ (n =-2, 2, 3, etc.). All these
  images have been calibrated following the procedure described in
  Section~\ref{sec:obs}. }
\label{fig:ima}
\end{figure*}

For the sources with peak flux densities above 6$\sigma$ and the
secure 5.4$\sigma$ detection (AMS12) cleaned images were produced and
 are shown in Figure~\ref{fig:ima}.  For these final images, the
tangent-point of the data was shifted to the centre of the detected
emission. The images were produced using a robust weighting of 5 and
the central region was cleaned. The properties of the radio
emission have been determined using the task {\tt IMFIT}.

In addition, on the 1st of November the source J171156.0+590639 was
also observed in order to test the observing strategy. This
source is located 1.38~deg away from the phase-reference calibrator
and had already been observed with NRAO's Very Large Baseline Array (VLBA) at 1.4
GHz. The flux density at 1.4~GHz from the VLBA of 17.03$\pm$1.29~mJy
(Wrobel et al. 2004) are, within the errors, consistent with the EVN
measurements at \cfreq\ of 14.9$\pm$4.4~mJy (32$\times$13~mas$^2$)
as shown in Figure~\ref{fig:ima}. The small difference in flux densities
is insignificant, and could be explained by the slightly
different observed frequencies provided that, at VLBI resolution,
J171156.0+590639 has a spectral index $\alpha\sim0.5$. The source at
EVN resolution is slightly elongated in the North-South direction
which is consistent with the VLBA observations of Wrobel et
al. (2004).

\section{Results}\label{sec:res}

The EVN observations result in 7 detections and 4 non-detections (AMS01,
AMS06, AMS16, and AMS17). Images for the detections are shown in
Figure~\ref{fig:ima}. Within the FoV mapped around each source
(4.1$\times$4.1 arcsec$^{2}$) no further radio emission has been
detected in any case.

Amongst the detected sources, the integrated fluxes are not
significantly higher than the peak fluxes (see Table~\ref{tab:obs}),
suggesting the recovered flux originates from an essentially
point-like region. The lack of other detections within the field of
view around each source rules out the detection of any double cores or knots separated by
more than $\sim$\size.

\begin{figure}
\includegraphics[angle=0,width=8.cm]{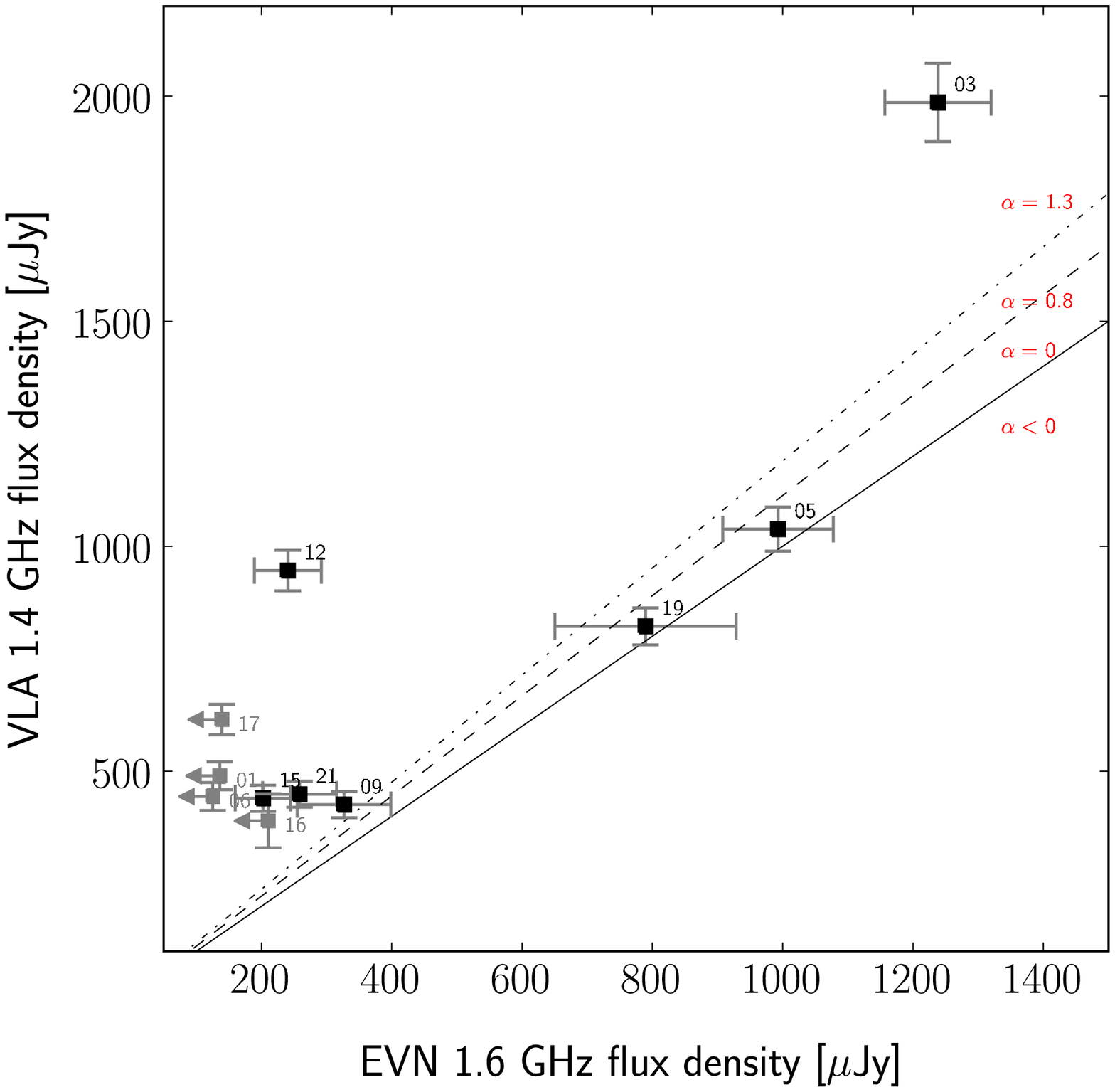}
\includegraphics[angle=0,width=8.cm]{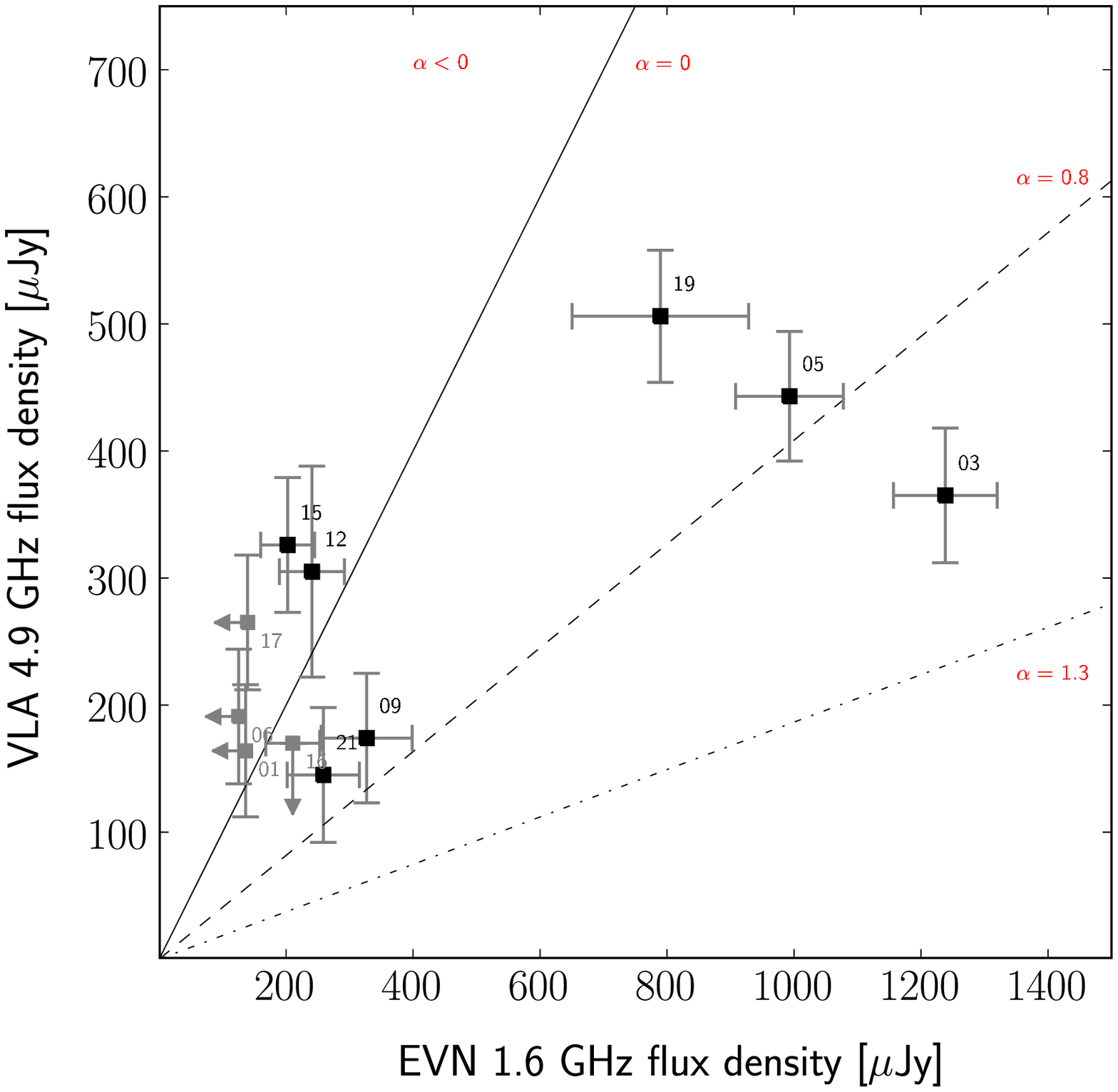}
%\vspace{11pc}
\caption{Radio flux densities at 1.4~GHz and 4.9~GHz (VLA-B and VLA-C; MS06b)
  versus the EVN flux densities at \cfreq . The solid line displays the track due to 
  a completely flat spectral index ($\alpha=$0.0), the dashed line displays a spectral index of
  0.8 and the dashed-doted one of 1.3. The flat spectral indices
  indicate optically-thick emission, most likely due to an AGN core
  given the $T_{\rm B}$ inferred. Extended emission from the lobes of
  jets and emission from star-formation typically lead to
  optically-thin emission with spectral index around $\alpha=$0.8
  (e.g. Condon 1992).  Inverse-Compton scattering against the Cosmic
  Microwave Background can lead to ultra-steep sources with
  $\alpha\sim1.3$ (e.g. Kardashev 1962; R\"ottgering et~al. 1994).}
\label{fig:evnvla}
\end{figure}

In Figure~\ref{fig:evnvla}, EVN and VLA flux densities are
displayed. In the top panel of Figure~\ref{fig:evnvla}, the clustering
of sources around low flux densities is a natural consequence of the
1.4~GHz flux limit used to select the sample.  In the bottom panel,
less clustering is visible, due to the range of different spectral
indices in the sample (see MS06b).

The flux densities at 4.9~GHz at the VLA resolution can be compared to
the EVN flux densities at 1.66~GHz, under the assumption that the
4.9~GHz flux emerges from the same physical region as the EVN cores
($\lesssim$\size). Given that the observed 4.9~GHz corresponds to
rest-frame between 14 and 25~GHz, this assumption is likely to be
appropriate (see e.g. Ulvestad et~al. 1999).  Under this assumption,
spectral indices have been determined for the cores (see
Table~\ref{tab:res}).  From Figure~\ref{fig:evnvla} and
Table~\ref{tab:res}, six sources show more flux at 4.9~GHz compared to
1.66~GHz, indicating an inverted core spectral index ($\alpha <0$). The
remaining sources show less flux at 4.9~GHz than at 1.66~GHz.
 
Using the spectral index $\alpha^{4.9}_{1.4}$ of Table~\ref{tab:obs},
flux densities at VLA resolution at the frequency of 1.66~GHz have
been derived ($S_{\rm VLA 1.6}$), to estimate the fraction of the
total flux density recovered at EVN resolution.  These quantities are
shown in Table~\ref{tab:res}. For the detected sources, the EVN
recovers between 29\% and 107\% of the low resolution flux densities,
and the median recovered fraction is 78\%. The median fraction
including non-detections is 64\%, so that typically two thirds of the
1.66~GHz radio flux density is confined to a region of $\lesssim$\size
. The percentages for the non-detected sources are $\lesssim$64\%
(although mostly $\lesssim$32\%) and for these sources, a compact
source with a fraction that is similar to that of some detections
(e.g. AMS12), cannot be ruled out. 

Table~\ref{tab:dirtso} also summarises the optical and low-resolution
radio properties of the sources. The recovered fraction for
blank-spectrum and narrow-line sources are statistically
indistinguishable: the probability of the two samples being drawn from
the same underlying distribution is 92\% using several two-sample
univariate tests (ASURV: Feigelson \& Nelson 1985). There is also no
correlation between the spectral indices $\alpha^{1.4}_{0.610}$,
$\alpha^{4.9}_{1.4}$ and the percentage of the recovered flux.  The
recovered fraction is apparently independent of the low-resolution
radio spectral properties. The characteristics of the detected and the
missing flux are discussed in Section~\ref{sec:dis}.

The luminosity densities $L_{\nu}$ at  the resolution of the EVN have been calculated using: 

\begin{equation}
\label{eq:lu}
L_{\nu_{\rm rest}} = { d_{\rm lum}^2  S_{ \nu_{obs} } \over   (1 + z) }  ,
\end{equation}

\noindent where $\nu_{\rm obs}$ is the observed frequency (1.66~GHz) and
$\nu_{\rm rest}=\nu_{\rm obs} \times(1+z)$ is the rest frame frequency, which
varies between objects in the sample due to their different redshifts.
$S_{\nu}$ is the flux density at 1.66~GHz and at EVN resolution while $d_{\rm
  lum}$ is the luminosity distance. Since at observed frequencies below
1.66~GHz the spectral indices are not known at mas resolution, the conversion
to rest-frame luminosity 1.66~GHz is not possible and no ``K-correction'' is
attempted.

Instead, the luminosities quoted in
Table~\ref{tab:res}, $L_{\nu_{\rm rest}} $, are at $\nu_{\rm rest} =
1.66\times(1+z_{\rm spec})$~GHz, where the $z_{\rm spec}$ is the spectroscopic
redshift for each source.

The brightness temperature $T_{\rm B}$ has been calculated using: 

\begin{equation}
\label{eq:tb}
T_{\rm B} = \frac{8\, {\rm log}(2)}{3 \pi} \frac{c^2}{2\,k} \frac{S_{{\nu}_{obs}} \,(1 + z)}{\theta_M \theta_m} ,
\end{equation}

\noindent where $c$ is the speed of light, $k$ is Boltzman's constant,
$\theta_M$ and $\theta_m$ are the semi-major and semi-minor angular sizes,
respectively (Condon et al. 1991).  Given that for the detected sources, the
median redshift is $z=2.5$, this corresponds to $\nu_{\rm
  rest}\sim5.8$~GHz (and to 8.6~GHz in the case of AMS16). The median
surface brightness temperature, $T_{B}$, is found to be $10^{6}$ K.

The detected and the missed emission have both been converted to a
star-formation rate (SFR), following Condon (1992). The SFRs derived
if the missing flux were due entirely to star formation are shown in
Table~\ref{tab:res}.  This is discussed further in
Section~\ref{sec:dis}.

\begin{table*}
\begin{center}
\begin{tabular}{llcccccccc}
  \hline
  \hline
  Name & $z$$^{a}$ & log$_{10}$($L_{\nu_{\rm rest}}$$^{b}$ & log$_{10}$($T_{\rm B}$$^{b}$ & 
  $S_{\rm VLA~1.6}$$^{c}$ & Fraction$^{d}$ & log$_{10}$($L_{\nu_{\rm miss}}$$^{e}$ &
  SFR$_{\rm miss}$$^{f}$ & $\alpha^{4.9}_{1.6}$$\,^{g}$& Diameter$^{h}$\\

 &  & {\tiny [\whzsr]})& {\tiny [K]}) & {\tiny [$\mu$Jy]} & {\tiny [\%]} & {\tiny [\whzsr]})
 & {\tiny [M\subsun yr$^{-1}$]} & {\tiny (core)} & {\tiny [pc]} \\
\hline
AMS01 & 2     & $<$23.0 & --     & 422       & $<$32       & $>$23.3 &  $>$1424 & $\leq$-0.17 $\pm$ 1.00 &  -- \\
AMS03 & 2.698 & 24.2    & 6.8    & 1579      & 78 $\pm$ 5  & 23.6    &  3249 & 1.13 $\pm$ 0.34     & 142 \\
AMS05 & 2.850 & 24.1    & 6.7    & 925       & 107 $\pm$ 3 & --      &  -- & 0.74 $\pm$ 0.30       & 140 \\
AMS06 & 1.8   & $<$22.9 & --     & 396       & $<$31       & $>$23.2 & $>$1066 & $\leq$-0.39 $\pm$ 0.94  & -- \\
AMS09 & 2.1   & 23.4    & 5.9    & 377       & 86 $\pm$ 1  & 22.6    &  279 & 0.58 $\pm$ 0.78      & 199 \\
AMS12 & 2.767 & 23.5    & 6.0    & 811       & 29 $\pm$ 1  & 23.9    &  5737 & -0.22 $\pm$ 0.73    & 149 \\
AMS15 & 2.1   & 23.2    & 5.5    & 422       & 47 $\pm$ 0  & 23.3    &  1218 & -0.44 $\pm$ 0.56    & 208 \\
AMS16 & 4.169 & $<$23.7    & 6.0    & $\leq$327 & $<$64  & $>$23.5    &  $>$2764 & $\sim$0.62          & -- \\
AMS17 & 3.137 & $<$23.4    & 5.8    & 548       & $<$25  & $>$23.8    &  $>$5363 & $\leq$-0.59 $\pm$ 0.70    & -- \\
AMS19 & 2.3   & 23.9    & 6.4    & 769       & 102 $\pm$ 2 & --      &  -- & 0.41 $\pm$ 0.43       & 172 \\
AMS21 & 1.8   & 23.2    & 5.9    & 385       & 67 $\pm$ 1  & 22.9    &  499 & 0.53 $\pm$ 0.91      & 152 \\
\hline
\hline

\end{tabular}

\caption{ \noindent Properties of the obscured
  quasars. $^{a}$Redshifts, are all spectroscopic except for
  AMS01. Redshifts with three decimal are from optical spectroscopy,
  while those with one decimal place are from mid-infrared
  spectroscopy (MS06a; MS08). The redshift for AMS05 is from Smith
  et al. (2009). $^{b}$Monochromatic EVN luminosity and brightness
  temperatures at restframe frequency $\nu_{\rm rest}=\nu_{\rm obs}
  \times(1+z)$. Note that the spectral index at small scales is not
  known and therefore no K-correction has been applied: both the the
  luminosity densities and the brightness temperatures are calculated
  in the rest frame using Eq.~\ref{eq:tb}. $^{c}$Expected total flux
  density (i.e. at VLA resolution) at the EVN observing frequency of
  1.66~GHz. This was estimated using radio spectral indicees based on
  the VLA measurements by MS06b.  $^{d}$Estimated fraction of total
  flux density at 1.66~GHz recovered at the resolution of the EVN, the
  quoted uncertainties do not include the 9\% uncertainty on the
  amplitude inferred from the phase callibrators.  $^{e}$Logarithm of
  the monochromatic restframe luminosity of the non-detected (missing)
  radio emission, calculated with Eq.~\ref{eq:lu}. The star-formation
  rate of the undetected radio emission.  $^{f}$This star-formation
  rate is calculated for massive stars only
  ($M_{\star}\geq5$~M$_{\odot}$). $^{g}$Core spectral index assuming
  that all the VLA emission emerges from the same physical region as
  the EVN emission.  $^{h}$Upper limits on the physical extend of the
  EVN emission region. The maximum of the deconvolved beam size is
  $\sim$17 mas, which translate at the median redshift of 2.3 to a
  physical extend of 137~pc (approx. 150~pc).}

\label{tab:res}
\end{center}
\end{table*}

\section{Discussion}\label{sec:dis}

Most of the observed sources show compact radio emission at physical
sizes $\lesssim$\size, and this emission accounts for a significant
fraction ( $\lesssim$30\% up to $\sim$100\%) of the total radio
emission. In the following discussion, the flux density detected by the EVN
observations is referred to as the ``core''.  For the missing flux,
resolved out by the EVN, there is no information about the structure
or extent. This radio emission could, in principle, emerge from
star-formation on either circum-nuclear or galactic scales, from the radio
emission due to a jet, or a combination of both. The two limiting
cases are considered here.

\subsection{The detected radio emission}\label{sec:det}

For the core emission in detected sources, the median brightness temperature,
$T_{B}=10^{6}$~K (see Table~\ref{tab:res}), is slightly high compared
to synchrotron radiation from supernovae (SNe) and SNe remnants only. The
regions with radio emission due to star-forming typically have $T_{\rm
  B}\lesssim10^{5}$~K \citep{mux94}, however the values of $T_{B}$ in this
sample are close to the limiting value. Therefore, the possibility that the
radio emission detected by the EVN are in fact dense star-forming regions
needs to be considered.

The detected cores have a median luminosity, $L_{\rm
  rest}=3\times10^{23}$~\whzsr. If this emision was due to
star-formation only, then the estimated star-formation rate (SFR),
following Condon (1992), would correspond to a SFR $\sim3600$ \msolyr\
(of massive stars only $\geq5$ M$_{\odot}$), and integrating over a
Salpeter (1955) initial mass function (IMF) would result in a median
total SFR of $\sim14~000$~\msolyr . The extremely high SFRs derived in
this way, and quoted in Table~\ref{tab:res}, are comparable or
generally greater than the SFRs inferred for submillimetre-selected
galaxies \citep{smail97}, except here they are confined to a
significantly smaller volume. Given that the resolution of 17~mas
corresponds to typically \size\, at $z=2.3$, these cores have physical
sizes comparable to those of the ``extreme starburst'' regions found by
Downes \& Solomon (1998), but the SFRs are about two orders of magnitude
higher.  These are unreasonable SFRs for such a small region.

The hypothesis of the detected cores being dense star-formation
regions can also be rejected on the basis of the mass of molecular gas
required to power such a SFR. An estimate of the typical
star-formation density of the cores, assuming a total SFR of 14~000
\msolyr\, in an area with diameter a 200~pc corresponds to
$\Sigma_{\rm SFR}$$=1.1\times10^{5}$ \msolyr\, kpc$^{-2}$.  Thus, if
these cores followed the Kennicutt-Schmidt law [${\Sigma_{SFR}}
= $${2.5 \times 10^{-4}}$${{ ({{\Sigma_{gas}} \over {1~M_{\odot}~{\rm
          pc}^{-2}}})}^{1.4}}$
M$_{\odot}$~yr$^{-1}$~kpc$^{-2}$, Kennicutt 1998], they would have
extreme gas densities of $\Sigma_{\rm gas}=$$1.5\times10^{6}$
\msol\,pc$^{-2}$, or a gas mass of $1.9\times10^{11}$ \msol\ (in a
characteristic size $\lesssim$\size). Such extreme molecular gas masses have been
observed around high-redshift quasars, but they are extended over
$\sim$kpc scales and not in such a small nuclear region (e.g. Walter
et al. 2004).  Thus, the hypothesis that the detected cores are
dominated by star-formation can be safely rejected, and in the
rest of this section we assume the cores to be due to an AGN.

Table~\ref{tab:res} quotes the core spectral indices,
$\alpha^{4.9}_{1.6}$, derived assuming that all the flux density at 4.9~GHz at
VLA resolution emerges from a region comparable to the EVN
resolution. Overall the spectral indices appear to be either
relatively steep, or inverted. However, we note that within the
uncertainties, most of the inverted spectra could be flat.

The flat or inverted spectrum can be understood as synchrotron
self-absorption due to a high density of relativistic electrons
causing the emission to be optically thick.  However, none of the
spectral indices are close to the expected value of $\alpha=-2.5$ for
an entirely self-absorbed regime (Rybicki and Lightman 1979).  Such
observed flat spectra can be observed if the transition between
optically-thin and optically-thick (and therefore the peak in the
spectrum) occurs at a frequency somewhere between 5.8 and 17~GHz. The
expected spectral indices would then be relatively flat, consistent
with those seen in Table~\ref{tab:res}. Alternatively, several
emitting components, each of which becomes self-absorbed at a
different frequency, when superimposed, can lead to the observed
spectral indices.

It is worth noting that the sources with high recovered fractions
(that is AMS03, AMS05, AMS09 and AMS19) have core spectral indices
indistinguishable from the low-resolution spectral indices. This
suggests the physical source of the EVN flux is also the dominant
component of the emission detected by the VLA and therefore apparently
compact. The actual mechanisms, however, probably vary from source to
source, as described below.

In the case of AMS05 and AMS16, the total (low-resolution) spectra
appear gigahertz-peaked (MS06b), reminiscent of the young compact powerful
radio sources (O'Dea 1998). The core spectral index of AMS05 is
consistent with the low-spatial resolution spectral index (between VLA
observations at 1.4 and 4.9~GHz), and 100\% of the low-resolution flux
density is recovered. This suggests the same emission is being seen at
VLA and EVN resolutions, so that the dominant source of radio emission
is compact ($\lesssim$\size). This also agrees with the hypothesis of
synchrotron self-absorption affecting frequencies around and below
5--6~GHz, while the emission at frequencies higher than
this is still optically-thin and therefore shows a steep spectrum. The
core spectral index of AMS16 is ill defined, since the source is not
detected at either 4.9 or 1.66~GHz, however a similar argument to that
of AMS05 would apply.

Most of the other sources have values of $\alpha^{4.9}_{1.6}$ much
flatter than their overall $\alpha^{4.9}_{1.4}$ value at VLA
resolution. This is consistent with the following (simplified)
scenario: flat spectrum emission from an optically-thick core contributing a
significant fraction of the total luminosity, and steep-spectrum
emission from optically-thin lobes from a jet.  At EVN resolution, the
emission from the jet lobes is resolved out, so that the core
dominates, and the spectral indices appear flat or inverted.

A slight modification is required in the case of AMS15 and AMS19:
their low-resolution spectral indices are also flat. This can be
explained by Doppler-boosting of the emission from the core, which
would make it a more important contributor to the low-resolution
spectrum, which would then appear overall flat. In the case of AMS19,
the core spectral index is also flat, and 100\% of the flux is
recovered, consistent with the beaming scenario. AMS15 has an inverted
core spectrum, and the EVN recovers only 47\% of the low-spatial
resolution flux density.

 Doppler-boosting is only expected when the radio jet is closely
aligned with the line-of-sight, so the jet must be close to face-on.  In the
unified scheme, only unobscured quasars can have such a configuration. AMS15
and AMS19 are obscured quasars with a face-on radio jet, and neither show
emission lines in their optical spectra.  These two objects are probably not
obscured by the torus of the unified scheme, but rather by dust on a larger
scale, presumably distributed along the host galaxy.  This provides further
evidence for the hypothesis of ``host-obscuration'' in some quasars (Mart\'\i
nez-Sansigre et al. 2005; MS06a; Rigby et al. 2006).

At first sight, one might expect flat-spectrum or gigahertz-peaked
sources to have the highest detection rates, and that is indeed the
case of AMS05 and AMS19.  However, some of the sources with the
highest recovered fraction (e.g. AMS09 and AMS21) have steep
spectra overall.  This can be reconciled if the total emission from AMS09 and
AMS21 is dominated by radio emission from the lobes of jets, but these
jets are small, with an extent limited to a few hundred pc. The high
recovered fractions (67 and 86\%) are caused by most of the lobe
emission being on scales smaller than the beam size, so that only a
small fraction of the emission is resolved out. The ``core'' spectral
indices of AMS09 and AMS21 are both $\sim$0.5, but within the large
uncertainties they agree with the total spectral indices (at low
resolution) between 1.4 and 4.9~GHz (compare their values in
Tables~\ref{tab:obs} and \ref{tab:res}).  This is in agreement with
the suggestion that the ``cores'' of these two sources are actually
unresolved lobes, and indeed the image of AMS09 hints at some extended
structure (Figure~\ref{fig:ima}). These sources appear to be
lower-luminosity
analogues to compact steep-spectrum sources (O'Dea 1998).\\

\subsection{The missing radio emission}\label{sec:miss}

In MS06a, the flux densities of this sample at 1.4~GHz obtained by the VLA
(5-arcsecond resolution, from Condon et al. 2003) and WSRT (14-arcsecond
resolution, from Morganti et al. 2004) were compared and found to be
consistent within the errors. There is no evidence for the VLA to `resolve out'
flux at 1.4~GHz on scales $\geq5$ and $\leq14$~arcseconds, and the the VLA
flux densities at 1.4~GHz (and $S_{\rm VLA 1.6}$) can be safely considered to
represent the entire flux densities of the radio sources at $\sim$GHz
frequencies.

There is no intermediate resolution data, and thus no information on scales
$\geq 17$ mas and $\leq 5$ arcseconds, and therefore the extent of the missing
radio emission correspond to $\grtsim$\size\, and $\lesssim$40~kpc.  Again
there is the possibility that the radio emission at intermediate size scales
is produced by star-formation, or jet activity or a combination of both.
Similar arguments for the core are used to discuss the nature of the radio
emission:

1. The entirety of the missing flux is due to star-formation. Following Condon
(1992), the estimated SFRs are in the range $\sim500-6000$ \msolyr (see
Table~\ref{tab:res}).  This is only for massive stars, with masses
$M_{\star}\geq5$~M$_{\odot}$, so the total star-formation rates would be
$\sim4\times$ greater if integrated over a Salpeter IMF. Thus, except in the
cases of AMS09 and AMS21 (and maybe AMS15), these are unphysically high SFRs.
This is not surprising, since this sample was selected using a radio criterion
to avoid contamination by luminous starbursts. Thus, in the sample as a
whole, it is extremely unlikely that the missing flux is all due to
star-formation.

2. If the missing flux is instead all due to AGN jets, then the median
jet luminosity is $L_{\nu_{\rm miss}} =2\times10^{23}$~\whzsr,
comparable to RQQ, RIQ, and FR~I radio galaxies.  Therefore, it is most
likely that the missing flux is dominated by emission from jets or
lobes on physical scales $\grtsim$\size, and $\lesssim$40~kpc.

For the sources AMS09, AMS15 and AMS21, the missing flux suggest
reasonable SFRs, indicating a significant contribution from
star-formation cannot be ruled out.  However, the cores of all 3
objects are clearly AGN dominated (see Section~\ref{sec:det}).  For
the rest of the sample, although the extended emission is likely to be
completely dominated by AGN emission, it is not possible to rule out
the additional presence and thus contribution of a powerful starburst.

\subsection{Comparing the core to the total  emission}

\begin{figure}
\hspace{-0.5cm}
\includegraphics[bb=25 145 560 670,clip,angle=0,width=9cm]{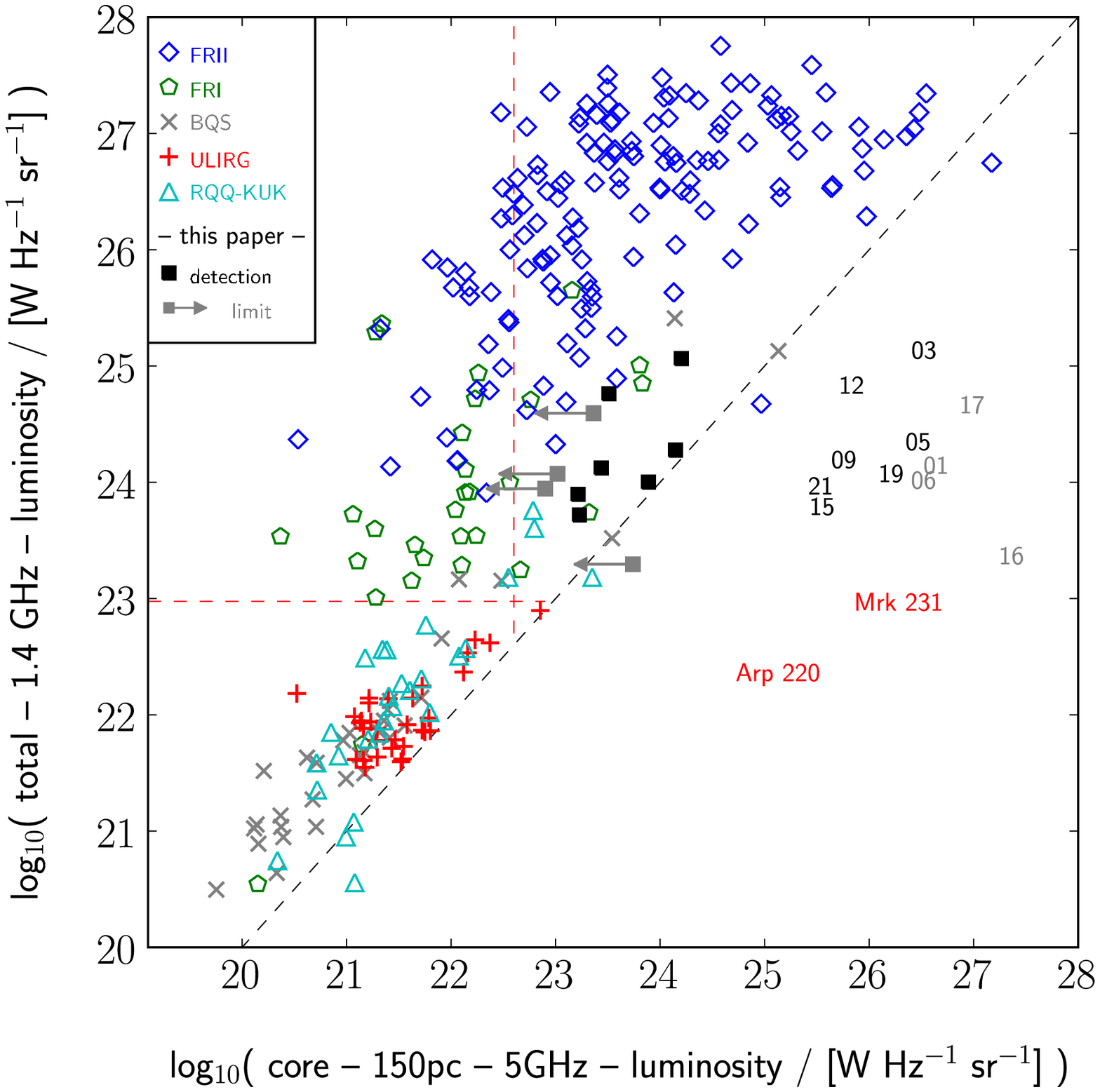}
\vspace{-0.5cm}
\caption{Total luminosity ($\nu_{\rm rest}\sim1.4$~GHz) versus core
  luminosity at $\nu_{\rm rest}\sim5$~GHz. The total luminosity
  densities of the sample of obscured quasars have been K-corrected
  with the spectral index $\alpha^{1.4}_{610}$. For the other samples
  an average spectral index of 0.8 has been assumed. To allow
  comparison, all the cores luminosity densities have been K-corrected
  using a spectral index of 0. The diagonal line indicates
  equality. The horizontal and vertical lines represent the luminosity
  for a SFR of 1000~\msolyr (all stars of 0.25 to 100~M$_{\sun}$, integrated over a Salpeter
  IMF), at 1.4 and 5~GHz, respectively.  The detected sources from
  this work are indicated by black rectangles while the grey
  rectangles and arrows show upper limits of the non-detections. In
  addition the total and core luminosities are shown for the FR~II and
  FR~I sources of the 3CRR catalogue (Blundell et al. in prep.), a
  sub-sample of the Bright Quasar survey (BQS, Miller et al. 1992;
  1993), Ultra-luminous infrared galaxies \citep[ULIRG, ][]{smith98} and
  a sample of radio-quiet quasars (Kukula et al. 1998).}
\label{fig:fr12ul}
\end{figure}

Based on the discussion in Sections~4.1 and 4.2, both the core and
most of the missing radio emission are dominated by AGN emission.  The
extended emission is unconstrained within the range $\grtsim$\size\,
and $\lesssim$40~kpc.  It is therefore both interesting and meaningful
to compare the radio properties of this sample to those of other
samples of AGN, particularly given that the total luminosities of the
sources are comparable with those of RIQ and FR~I, and the core
luminosities of the observed sources are also similar to the core
luminosities of local FR~I, which range between 10$^{22}$ --10$^{24}$
\whzsr\, \citep[e.g.][]{zirbel95}.

Figure~\ref{fig:fr12ul} shows the total luminosity density versus
core luminosity density for other samples of AGN: FR~I and FR~II
sources from the 3CRR sources (Blundell et al. in prep.), a sub-sample
of the Bright Quasar survey (BQS, Miller et~al. 1992; 1993), ULIRGs
\citep{smith98} and a sample of radio-quiet quasars (Kukula et
al. 1998, actually a subsample of the Miller et al. 1993 sample).

The obscured quasars observed in this paper show a narrow distribution
of the ratio of their total to core luminosities, i.e. these
luminosities being correlated, with the null hypothesis having a
probability 0.02\% (ASURV software using the sub-routine
BIVAR). The total luminosities are therefore related by a linear
regression with log($L_{\rm tot}$) $\sim$ (3.5$\pm$1.9) log($L_{\rm
  core}$). The sample of RQQ shows a correlation significant at the
$\geq3\sigma$ level with log($L_{\rm tot}$) $\sim$ (1.0$\pm$0.1)
log($L_{\rm core}$), the FR~I sources show a slope of 0.7$\pm$0.2
(97\% confidence) while the FR~II sources have 0.4$\pm$0.1
($\geq3\sigma$ significance). Hence, as well as having similar
recovered fractions to the RQQ, there are hints that the obscured
quasars show a slope more similar to the RQQ than the FR~I or FR~II
sources. Overall, the obscured quasars appear more similar to the RQQ
than any of the FR~I or FR~II radio sources, although the errors in
the small sample of objects observed in this paper are high.

The observations are at a range of different frequencies, which have
been converted to approximately 5~GHz (rest-frame). The luminosities
of obscured quasars are at $\nu_{\rm rest}=\nu_{\rm obs} \times(1+z)$,
corresponding to around 5.8~GHz. This small difference is not
important: the main result of Figure~\ref{fig:fr12ul} is that the
sample of obscured quasars shows similar core-to-total luminosity
ratios to the radio-quiet and radio-intermediate quasars (as well as,
for example, to Mrk~231; Ulvestad et al. 1999; Kl\"ockner et
al. 2003).  In these samples, the core emission reprents a large
fraction of the total emission, whereas in FR~I and FR~II sources, the
total emission is typically $\geq$1~dex greater than the core
emission.

In the sample of obscured quasars, typically $\sim$60\% of the flux is
recovered on scales $\lesssim$\size, and both flat-spectrum or
inverted cores are inferred.  The distributions of recovered fraction
from the sample presented here and from the RQQ sample of Kukula
et~al. (1998) are indistinguishable.  For many of the objects the
physical scales are only constrained to be $\lesssim$1~kpc.  However,
amongst their lowest-redshift sources ($z\sim0.05$) the VLA-A array
resolution at 8.4~GHz is 0.24~arcseconds, which corresponds to
$\sim$230~pc scales, and again the distribution of recovered fractions
is similar to that of the $z\grtsim2$ obscured quasars.  The main difference is
that the typical spectral index inferred for the cores of Kukula
et~al. is $\sim$0.7.  About half of the sample presented here has core
spectral indices consistent with those of Kukula et al. (1998), while
the other half have inverted (or, given the errors, possibly flat)
spectra.

The RQQ and Seyferts show a correlation between $M_{\rm V}$ and
$L_{\rm 8.4~GHz}$. The obscured quasars have mid-infrared luminosities
that correspond to unobscured quasars with estimated values of $-26
\lesssim M_{\rm B} \lesssim -24$ (see MS06a and Smith et~al. 2009),
and radio luminosities at $\sim$6~GHz of $10^{23}-10^{24}$ W
Hz$^{-1}$ sr$^{-1}$. Assuming $M_{\rm B} \approx M_{\rm V}$ and $L_{\rm
  8.4~GHz}\approx L_{\rm 6~GHz}$, for a given value of $M_{\rm V}$,
the obscured quasars are $\sim$2 dex more powerful in radio
luminosity compared to the RQQs. This fits in with the results of
Figure~\ref{fig:fr12ul}: the high-redshift obscured quasars have
compact-to-total ratios similar to those of RQQ, and extended emission
on comparable scales, yet the RQQ are significantly less powerful at
radio frequencies.

Finally, given that the obscured quasars have radio luminosities corresponding
to RIQ,  and that $\sim$40\% of the flux density is in an extended component,
it is clear that the majority of them are not beamed, and that their
luminosity densities are intrinsic.

\section{Conclusions and summary}\label{sec:conc}

A subsample of 11 high-redshift obscured quasars has been observed at
1.66~GHz with the EVN, reaching a noise level of $\sim$~\rms. The
observations are sensitive to radio emission on physical scales
$\lesssim$\size. Seven sources are securely detected and show compact
radio emission.

Amongst the detections, $\sim$30--100\% of the entire flux density is
recovered in the core. Considering the luminosity densities and
brightness temperatures, and comparing them to SNe brightness
temperatures as well as SFRs and molecular gas masses of extreme
starbursts, the cores are found to be AGN-dominated and not due to a
nuclear starburst. In the non-detections, the limits inferred for the
radio emission and inferred properties are similar to those of the
detected cores.

Comparing the total and core flux densities and spectral indices, the
sample is found to contain a mixture of steep-spectrum,
gigahertz-peaked spectrum, compact steep spectrum and beamed sources.

When considering the EVN observations, no correlations are found
between the optical spectra and the recovered fraction. Hence, there
is no difference in the recovered fraction between host-obscured or
torus-obscured quasars. Indeed, no correlation is found either between
the low-resolution spectral indices and the recovered fraction. This
can be explained by the sample being composed of a mixture of
steep-spectrum, compact-steep-spectrum, gigahertz-peaked and
flat-spectrum sources. Hence, the recovered fraction is related to the
size and orientation of the jet rather than the distribution of the
obscuring dust.

However, when considering the total (low-resolution) spectral indices and
the core spectral indices, two sources are consistent with overall flat
spectra due to a Doppler-boosted radio core (from a face-on jet, AMS15 and
AMS19). In the unified scheme, only unobscured quasars are expected to show
such flat spectra, but neither source shows either broad or narrow emission
lines, despite being at redshifts where the rest-frame ultraviolet lines are
observable. These two sources are candidates for ``host-obscured'' quasars,
where the obscuring dust is distributed on $\sim$kpc scales, rather than in
the torus of the unified scheme.

At the redshifts of the sources, the missing flux densities (the
emission `resolved-out' by the EVN) correspond to too high radio
luminosities to be consistent with star-formation only in all but a
few cases. The extended emission is therefore also interpreted as
being dominated by radio emission from the lobes of jets, comparable
in luminosity to those of RIQ and FR~I radio sources. However, a
hybrid scenario with significant on-going star-formation cannot be
ruled out.

Since the core and the extended emission are both AGN-dominated, the
core and total luminosities of the sources are compared to those of
other samples of AGN: RQQ, RIQ, RLQ, FR~I and FR~IIs.  In particular,
the total radio luminosities of the obscured quasars are comparable to
those of FR~I and RIQ.  The core-to-total luminosity ratios
($\sim$0.5) are found to be more similar to those of RQQ, rather than
FR~I sources. Indeed, the RQQ studied by Kukula et~al. (1998) had a
mean core-to-total luminosity fraction of 0.6 at 4.8~GHz.  Although,
the cores of their RQQ had typically steep spectral indices, whereas
the high-redshift obscured quasars sometimes have inverted or flat
cores. However, it is worth noticing that it is very difficult to
assess the precise differences between RQQ and FR~Is when they have
been observed with different interferometers and at different
redshifts; inverse-compton scattering against the CMB could extinguish
old extended radio jets on larger scales in high redshift objects
(e.g. Blundell and Rawlings 2001).

The sample of obscured quasars have higher values of $L_{\rm radio}$
for an estimated $M_{\rm V}$, situating them in the RIQ regime rather
than the RQQ. Yet, their properties are more similar to those of RQQ
than to FR~I sources. These obscured RIQ have overall (low-spatial
resolution) steep spectra, and $\sim$40\% of the flux is resolved out
on $\grtsim$\size\, scales, so that they are not all beamed RQQ: they
have intrinsically higher luminosities. The sources in our sample
therefore reflect a gradual transition of intrinsic luminosities
between the values observed for RLQ and for RQQ.

\section*{Acknowledgments}

The European VLBI Network is a joint facility of European, Chinese,
South African and other radio astronomy institutes funded by their
national research councils.
ParselTongue was developed in the context of the ALBUS project, which
has benefited from research funding from the European Community's
sixth Framework Programme under RadioNet R113CT 2003 5058187. 
We thank Paul Alexander, Dave Green, and Julia Riley for comments on
the original observing proposal for this project.

\end{document}